\documentstyle[osa,epsfig]{revtex}  

\begin{document} 
\title{Three-dimensional spontaneous magnetic reconnection in neutral current
       sheets} 
\author{J\"org Schumacher$^1$, Bernhard Kliem$^2$, and 
        Norbert Seehafer$^3$}            
\address{$^1$Fachbereich Physik, Philipps Universit\"at Marburg, 
             D-35032 Marburg,  Germany}
\address{$^2$Astrophysikalisches Institut Potsdam, D-14482 Potsdam, Germany}
\address{$^3$Institut f\"ur Physik, Universit\"at Potsdam, D-14415 Potsdam, 
             Germany}
\maketitle

\centerline{\small\sl Phys. Plasmas, Vol.\ 7, No.\ 1, January 2000 issue}

\begin{abstract}
Magnetic reconnection in an antiparallel uniform Harris current sheet
equilibrium, which is initially perturbed by a region of enhanced resistivity
limited in all three dimensions, is investigated through compressible
magnetohydrodynamic simulations.  Variable resistivity, coupled to the
dynamics of the plasma by an electron-ion drift velocity criterion, is used
during the evolution.  A phase of magnetic reconnection amplifying with time
and leading to eruptive energy release is triggered only if the initial
perturbation is strongly elongated in the direction of current flow or if the
threshold for the onset of anomalous resistivity is significantly lower than
in the corresponding two-dimensional case.  A Petschek-like configuration is
then built up for $\sim10^2$ Alfv\'en times, but remains localized in the
third dimension.  Subsequently, a change of topology to an O-line at the
center of the system (``secondary tearing'') occurs.  This leads to enhanced
and time-variable reconnection, to a second pair of outflow jets directed
along the O-line, and to expansion of the reconnection process into the third
dimension.  High parallel current density components are created mainly near
the region of enhanced resistivity.

\end{abstract}


\section{INTRODUCTION}

The reconnection of magnetic field lines is one of the fundamental processes
that lead to the eruptive release of stored magnetic energy in plasmas.  A
large number of observations and experiments, e.g., solar
flares,\cite{MAS94,TSU96} geomagnetic substorms,\cite{BA96} the sawtooth
instability in fusion experiments,\cite{ED86} and laboratory reconnection
experiments,\cite{YA90,YA97} support this hypothesis.  Magnetic reconnection
is important for the evolution of magnetohydrodynamic (MHD) instabilities
(e.g., the tearing mode\cite{FKR63} and magnetic island
coalescence\cite{PW79,BW80,SK97}) in current sheets which are generally
involved in the energy release process.  It is also important at the
dissipative scales in MHD turbulence.\cite{BW89,PO89} To explain the short
time scales of the energy release processes in spite of large values of the
Lundquist number $S=\mu_0LV_A/\eta_0$, fast reconnection models have been
proposed but are generally restricted to two-dimensional geometry, including
the stationary Petschek model\cite{PE64} and the spontaneous fast reconnection
model.\cite{UG84} Both these models are one-fluid descriptions in which
resistivity provides the nonideal effect required for reconnection and in both
models the resistivity is localized at the central magnetic X-point.

Models distinguishing between electron and ion effects, such as Hall
MHD,\cite{HU95,LS97} two-fluid MHD,\cite{BSD97} or kinetic
treatments,\cite{PR94,HE99} permit still higher reconnection rates, even in
the absence of resistivity.  However, all these models require significant
computational effort, which still prevents following the evolution of
microscopic perturbations to large spatial and temporal scales in three
dimensions.

In the spontaneous fast reconnection model, a current sheet equilibrium is
initially perturbed by a localized region of enhanced (anomalous) resistivity
in the sheet center, which leads to the development of a simple X-point
structure.  Anomalous resistivity is permitted to occur in this model also at
later times of the evolution if a threshold of the local current density 
${\bf j}$ or of the local electron-ion drift velocity 
${\bf v}_D=(m_i/e)\,{\bf j}/\rho$ is exceeded ($m_i$ -- ion mass, $e$ --
elementary charge, $\rho$ -- mass density).  This couples the
reconnection-driven flow to the evolution of the resistivity and enables a
positive feedback that amplifies the dynamics in the reconnection region.
This mechanism is of particular interest for eruptive energy release events
since it involves a threshold for onset, is initially self-amplifying, and is
fast, i.e., it leads to rates of magnetic flux change at the magnetic null
point (reconnection rates) higher than the Sweet-Parker rate\cite{SW58,PAR63}
and to Alfv\'enic outflows.

Several basic 3D effects of magnetic reconnection have already been analyzed
in simulations using temporarily or spatially constant resistivity such as the
diversion of current flow by the reconnection-driven
outflow,\cite{BI81,SCH91A,SCH91B,BH96} the creation of interlinked magnetic
flux tubes by multiple areas of enhanced resistivity,\cite{OT95} and the
tunneling of magnetic flux tubes.\cite{DA97} The very question of the
formation and maintenance of the spontaneous fast reconnection process in a
three-dimensional system with dynamically coupled resistivity has only
recently been investigated by Ugai and Shimizu.\cite{US96} The latter authors
generalized the 2D model by introducing a three-dimensional initial
perturbation by anomalous resistivity $\eta_{an}(x,y,z,t)$ in a triple current
sheet geometry with antiparallel magnetic field, ${\bf{B}}=B(y)\,{\bf{e}}_x$.
By varying the $z$ extension of the initial perturbation, they found that the
spontaneous fast reconnection process occurs also in 3D but only if the region
of perturbation is strongly anisotropic, extending in the $z$ direction by at
least 4 current sheet widths.

Such an anisotropy with preferred elongation {\em across} the magnetic field
cannot be expected to occur spontaneously without additional assumptions.  In
magnetized hot plasmas where the mean free path exceeds the ion cyclotron
radius, $\lambda_{mfp}\gg r_{ci}$, spatial scales along the field are
generally much larger than the scales across the field.  The results of Ugai
and Shimizu therefore cast doubt on the usefulness of the spontaneous
reconnection model to explain large-scale energy release events, such as solar
flares or magnetospheric substorms.  In this paper we extend their study and
find parameter settings that permit us to relax the requirement on the
anisotropy of the initial perturbation.

Ugai and Shimizu further suggested that a quasi-stationary regime of fast
reconnection, similar in structure to Petschek's model, is reached after
several $10^1$ Alfv\'en times ($\tau_A$) if the initial perturbation is
chosen to be sufficiently anisotropic (so that the 3D reconnection process
evolves in a manner similar to the well-known 2D behavior).  This aspect is
also important with regard to astrophysical applications because the energy
release events typically last several orders of magnitude longer than the
period of time that can be followed in a numerical experiment.  Whether
stationary Petschek-like reconnection is possible is still a matter of
debate.\cite{KU98} The observations of solar flares, for example, suggest that
the energy release is typically highly variable in time.\cite{AS95}
Two-dimensional numerical experiments on reconnection have shown that the
process of ``secondary tearing,'' which changes the topology to form a central
O-point and possibly a sequence of new X-points, accompanied by highly
variable reconnection rates, can be important at
$\agt10^2\,\tau_A$.\cite{SR87} We have therefore integrated the equations for
several $10^2\,\tau_A$ to study at least the beginning of the long-term
evolution of the three-dimensional system and to see whether a (quasi-)
stationary state is approached.

Furthermore, the ``standard'' equilibrium, the Harris current sheet, is used
in this paper without additional current sheets at the upper and lower
boundaries of the box (which slows down the initial evolution of the system
somewhat in comparison to the equilibrium employed by Ugai and Shimizu).  The
investigation is also restricted to an initially antiparallel magnetic field
configuration (a neutral current sheet), for which the comparison to the
two-dimensional case is most direct.  The presence of a magnetic guide field
component, $B_z\ne0$, leads in a three-dimensional system to a fundamentally
different topology by breaking the symmetry with respect to the midplanes of
the system and will be considered in a separate paper.

The outline of the paper is as follows.  In Sec.~\ref{model} we present the
equations and the numerical method.  In the next section we describe the
qualitatively new properties of the three-dimensional dynamics and compare
them with the two-dimensional case.  The influence of several parameters, such
as the anisotropy of the initial perturbation, the threshold for the onset of
anomalous resistivity, and the plasma beta on the build-up of fast
reconnection is considered.  A study of the long-term evolution
($>10^2\,\tau_A$) of the reconnection rate under the influence of secondary
tearing in 3D is performed for the first time.  We discuss the occurrence of
parallel currents and relate it to the current helicity density to explain new
findings that extend previous results.\cite{BI81,SCH91A,SCH91B,BH96}
In Sec.~\ref{conclusions} we give the conclusions and a discussion of the
applicability to astrophysical phenomena, such as solar flares.

\section{SIMULATION MODEL}
\label{model}

The compressible MHD equations are employed in the following form:
\begin{eqnarray}
\partial_{t}\rho&=&
                 -\nabla\cdot(\rho\,{\bf u}),               \label{eq_rho}\\
\rho\,\partial_{t}{\bf u}&=&
                 -\rho\,(\,{\bf u}\cdot\nabla\,)\,{\bf u}
                 -\nabla\,p+{\bf j}\mbox{\boldmath$\times$}{\bf B},\,\,\,\,\\
\partial_{t}{\bf B}&=& 
    \nabla\mbox{\boldmath$\times$}(\,{\bf u\mbox{\boldmath$\times$} B}\,)
   -\nabla\mbox{\boldmath$\times$}(\,\eta\,{\bf j}\,),\\
\partial_{t}U&=&
              -\nabla\cdot{\bf S},                         \label{eq_energy}
\end{eqnarray}
\noindent
where the current density {\bf j}, the total energy density $U$, and the flux
vector ${\bf S}$ are given by 
\begin{eqnarray}
{\bf j}&=& \frac{1}{\mu_0}\,\nabla\mbox{\boldmath$\times$}{\bf B},   
\nonumber\\
U      &=& \rho\,w+\frac{\rho}{2}\,u^{2}+\frac{B^{2}}{2\,\mu_0},     
\nonumber\\
{\bf S}&=&(\,U+p+\frac{B^{2}}{2\,\mu_0}\,){\bf u}
         -({\bf u}\cdot{\bf B})\frac{{\bf B}}{\mu_0} 
         +\eta{\bf j}\mbox{\boldmath$\times$}\frac{{\bf B}}{\mu_0}, 
\nonumber
\end{eqnarray}
and $w$ is the internal energy per unit mass, which is related to the pressure
through the equation of state, $p=(\,\gamma-1\,)\,\rho\,w$. The ratio of 
specific heats is $\gamma=\frac{5}{3}$.  The electric field is given by 
\begin{eqnarray}
{\bf E}=-{\bf u}\mbox{\boldmath$\times$}{\bf B}+\eta\,{\bf j}.
\label{Ohm}
\end{eqnarray}

An antiparallel Harris equilibrium (a neutral current sheet) with uniform 
density is chosen as the initial condition:
\begin{eqnarray} 
B_x &=&-\,B_0\,\tanh(y/l_{CS}),                              \label{eq1}\\ 
B_y &=&B_z=0,        \\ 
u_x &=&u_y=u_z=0,    \\
\rho&=&\rho_0,       \\
p   &=&(1+\beta)B_0^2/(2\mu_0)-B_x^2/(2\mu_0),               \label{pre}
\end{eqnarray}
where the plasma beta is defined as $\beta=2\mu_0\,p(|y|\to\infty)/B_0^2$. 

The variables are normalized by quantities derived from the current sheet half
width $l_{CS}$ and the asymptotic ($|y|\to\infty$) Alfv\'en velocity
$V_A=B_0/(\mu_0\rho_0)^{1/2}$ of the configuration at $t=0$.  Time is measured
in units of the Alfv\'en time $\tau_A=l_{CS}/V_A$, and $p$, {\bf E}, {\bf j},
and $\eta$ are normalized by $B_0^2/(2\mu_0)$, $V_AB_0$, $B_0/(\mu_0l_{CS})$,
and $\mu_0l_{CS}V_A$, respectively.  The normalized variables are used
henceforth.

The magnetostatic equilibrium is initially perturbed by a three-dimensional
anomalous resistivity profile.  An ellipsoid of enhanced resistivity is put
into the simulation domain at the origin for $0\le t\le t_0$, similar to the 
model used by Ugai and Shimizu\cite{US96}: 
\begin{eqnarray} 
\eta({\bf x})\!=\!C_1\exp[-(x/l_x)^2-(y/l_y)^2-(z/l_z)^2]\,.
\end{eqnarray}
In all simulations $t_0\!=\!10$, $C_1\!=\!0.02$, and $l_y=0.8$.  The
elongation of the initial perturbation in the $z$ direction is varied in the
range $l_z\in[0.8,8]$.  Also the elongation $l_x$ is varied in a few runs from
the standard value $l_x=0.8$.  For $t>t_0$ the resistivity is determined
self-consistently from the local value of the relative electron-ion drift
velocity ${\bf v}_D=(m_i/eB_0\tau_A)\,{\bf j}/\rho$ for every time step:  an
``anomalous'' value, $\eta_{an}$, is set if a threshold, $v_{cr}$, is
exceeded.  The resistivity model for $t>t_0$ is given by
\begin{equation} 
\eta({\bf x},t)=\left\{
  \begin{array}{l@{\quad:\quad}l}       0           & |{\bf v}_D|\le v_{cr}\\ 
    \begin{displaystyle}
     C_2\,\frac{(|{\bf v}_D({\bf x},t)|-v_{cr})}{v_0}
    \end{displaystyle}                              & |{\bf v}_D| >  v_{cr}. 
  \end{array}
 \right.  
\label{etaandef} 
\end{equation}
Here $C_2\!=\!0.003$, and the threshold value $v_{cr}$ is varied in the 
range $v_{cr}=(1.1\mbox{--}3)\,v_0$, where $v_0=|{\bf v}_D(0)|$.  

The choice of $v_{cr}$ implies an assumption on the initial current sheet half
width:  $l_{CS}(t\!=\!0)=(v_{cr}/v_0)\,l_{cr}$, where $l_{cr}$ is the critical
current sheet half width.  Using the ion thermal velocity as an
order-of-magnitude estimate of the critical electron-ion drift velocity for
the onset of kinetic current-driven instabilities and the excitation of
anomalous resistivity, $v_{cr}\sim(T_i/m_i)^{1/2}$, Ampere's law yields
$l_{cr}\sim4\beta^{-1}r_{ci}$, where $r_{ci}$ is the ion cyclotron radius.
The critical current sheet half width can be much larger than the scale
lengths at which the Hall and the pressure gradient terms in the generalized
Ohm's law become important.  These are, respectively, the ion inertial length,
$d_i=c/\omega_{pi}$, and $d_i\beta/4$.  The above estimate yields
$l_{cr}\sim2\beta^{-1/2}d_i$.  Thus, in the range of beta values where
magnetic reconnection is energetically important, $\beta\alt1$, one can expect
that the excitation of anomalous resistivity plays a role during the formation
of thin current sheets ($l_{CS}\sim d_i$) at the same time as, or even before,
the Hall and pressure gradient terms become significant.

Equations (\ref{eq_rho})-(\ref{eq_energy}) are integrated on a non-equidistant
Cartesian grid with uniform resolution in a range around the origin and slowly
degrading resolution toward the outer boundaries (see Table~{\ref{Tab1}).  A
two-step Lax-Wendroff scheme\cite{UG84,SK96} is used.  To ensure numerical
stability, artificial smoothing is applied to all variables integrated by the
scheme after each full time step in the following manner:
\begin{eqnarray*}
{\bf\Psi}_{ijk}^{n}&\longrightarrow&\sigma{\bf\Psi}_{ijk}^{n}
   +\frac{1-\sigma}{6}\,\times\\
   & &({\bf\Psi}_{(i+1)jk}^{n}+{\bf\Psi}_{(i-1)jk}^{n}\\
   &+& {\bf\Psi}_{i(j+1)k}^{n}+{\bf\Psi}_{i(j-1)k}^{n}\\
   &+& {\bf\Psi}_{ij(k-1)}^{n}+{\bf\Psi}_{ij(k+1)}^{n}). 
\end{eqnarray*}
The smoothing parameter $\sigma$ is set in the range [0.97,0.99], as
required by the different runs.  A variable time step is chosen to limit the
numerical diffusion\cite{SK96} and to satisfy the Courant criterion.

Since both the equilibrium and the initial perturbation possess spatial
symmetry and $B_z(t\!=\!0)=0$, all variables remain even or odd with respect
to the midplanes throughout the evolution, and the simulation domain can be
restricted to one octant of the cube
$[-L_x,L_x]\!\times\![-L_y,L_y]\!\times\![-L_z,L_z]$.  The boundary conditions
are thus symmetric at the planes $x=0$ (where $B_y$, $B_z$, $u_x$ are odd),
$y=0$ (where $B_x$, $B_z$, $u_y$ are odd), and $z=0$ (where $B_z$, $u_z$ are
odd), with even mirroring of the other integration variables.  Open boundary
conditions are realized at the other three boundary planes by the requirement
that the normal derivatives of all variables vanish, except for the normal
component of ${\bf B}$, which is determined from the solenoidal condition.

\section{RESULTS}

\subsection{Initial evolution}

The first two phases in the evolution of the perturbed current sheet are
largely similar to the two-dimensional case.  The Ohmic dissipation by the
localized initial resistivity causes a short ($0<t\alt2\tau_A$), transitory
phase of expansion of the heated plasma into all directions.  This is soon
terminated because the main effect of the perturbation is the reconnection of
magnetic field lines and the resulting $B_y$ component leads to predominant
acceleration of plasma out of the perturbed region by the Lorentz force
$f_x\approx-j_zB_y$.  The pressure then starts to decrease locally (already at
$t\approx1\tau_A$), which reverses the initial expansion in $y$ and $z$ and
causes acceleration of plasma in this second phase into the area of the
perturbation.  In a 2D system the inflow is in the $y$ direction only,
carrying new magnetic flux into the region.  This enables continuation of the
reconnection process and, in later phases, its amplification by re-creating
high current densities and the corresponding anomalous resistivity.\cite{UG84}
In the 3D system, however, the inflow is driven by forces in the $y$ and $z$
directions
[$F_{y,z}=(-{\nabla}p/2+{\bf{j}}\mbox{\boldmath$\times$}{\bf{B}})_{y,z}$].
The inflow in the $z$ direction generally acts to reduce the inflow in the $y$
direction necessary for ongoing reconnection.  For small $|F_z/F_y|$, the fast
reconnection regime is set up, similar to the 2D case, in a slab around the
plane $z=0$.  Otherwise the dominant inflow $u_z$ strongly reduces the inflow
$u_y$ and, correspondingly, the reconnection during the second phase.  This
leads to a weakening of the driving outflow $u_x$.  Subsequently the inertia
of the $u_z$ flow causes a pressure increase in the center of the perturbed
region and a second reversal of $u_y$, and the reconnection ceases.  The
magnetic X-line configuration then either diffuses away or undergoes a
transition to an O-line in place of the former X-line as a result of the
reversed $u_y$ flow.  Clearly, $l_z$ is the parameter that controls the ratio
of forces, with $|F_z/F_y|\to0$ for $l_z\to\infty$.  Since the development of
fast reconnection does not require a vanishing $F_z$, there is a finite value
$l_z^*$ above which fast reconnection occurs in a 3D system.

It might be expected that $l_z^*$ is rather large, since the acceleration of
the fluid in the $z$ direction is qualitatively different from the
acceleration in the $y$ direction.  Along the $z$ axis, the Lorentz force
vanishes due to the symmetry of the considered system, and only the pressure
gradient $-\partial_zp/2$ can accelerate the fluid.  In the $y$ direction, the
acceleration is determined by the balance between the pressure gradient
$-\partial_yp/2$, which remains directed outward as in the initial
equilibrium, and the inward-directed Lorentz force $f_y\approx j_zB_x$.  Both
quantities are tightly coupled during the whole evolution with their sum being
much smaller than each individual component.  Therefore, dominance of the
inflow $u_z$ might be expected for
$|\partial_zp/2|\agt|j_zB_x-\partial_yp/2|$, i.e., values of $l_z$ of the same
order as $l_x$ and $l_y$.  This is in line with the conclusions by Ugai and
Shimizu.\cite{US96} However, it will be seen below that $l_z^*$ depends
strongly on other parameters in the system and is not necessarily large.

\subsection{Build-up of fast reconnection}
\label{buildup}

Figure~\ref{comp} shows the evolution of the Harris sheet at characteristic
times for $l_z=1$ and $l_z=8$, using $\beta=0.15$ and $v_{cr}=3$.  In the
system with $l_z=1$ (run~1 in Table~\ref{Tab1}) the flow $u_y$ is already
reversed (directed away from the reconnection region) by the inertia of the
$u_z$ inflow and reconnection has already stopped by $t=24$.  Also the field
lines in the plane $z=0$ no longer possess the pure X-type topology,
characteristic of reconnection, which was enforced during the initial
perturbation ($t<t_0=10$).  No anomalous resistivity occurs in this run after
$t=t_0$, and characteristic quantities, such as the outflow velocity and the
current density, decrease subsequently.  In contrast, for $l_z=8$ (run~2 shown
at $t=48$) the field lines and the flow in the plane $z=0$ are identical to
the picture known from 2D systems after those have evolved into the fast
reconnection regime, where anomalous resistivity is again excited.  Also the
current density ridges near the separatrices of the magnetic field are clearly
developed, with the maxima of $j=|{\bf{j}}|$ and $\eta_{an}$ lying at
${\bf{x}}=0$.  The maximum outflow velocity $\max(u_x)$ rises continuously
until it becomes Alfv\'enic at $t\!\approx\!160$ --- before the boundary at
$|x|=90$ is reached.

Runs~1 and 2 are in general agreement with the results of Ugai and
Shimizu.\cite{US96} The removal of the pair of oppositely directed current
sheets at the upper and lower boundary of the numerical box and the more
gentle profile of the current density in the Harris equilibrium used here
raise the minimum elongation of the initial resistivity, for the development
of fast reconnection, somewhat to a value $l_z^*\sim8$.  The need to search
for parameter settings that render such a strong requirement unnecessary is
thus even strengthened.

Simply enlarging the $x$ extent of the perturbation ($l_x$) does not support
the development of fast reconnection since the initial configuration is then
closer to the Sweet-Parker configuration.  This possesses smaller reconnection
rates than a configuration with more localized resistivity (runs~4 and 5 in
Table~\ref{Tab1}; see also Sec.~\ref{long}).  Applying the initial
perturbation for a longer time appears to be reasonable because the
collisional time scale, at which spontaneously created anomalous resistivity
is expected to decay, can be much longer than the Alfv\'en transit time
$\tau_A$ of the thin sheets forming in cosmic plasmas.  However, this does not
change the result for $l_z=1$:  if the initial perturbation is applied for a
longer time (e.g., $t_0=30$), the reversal of the flow $u_y$ and a transition
from X-type to O-type magnetic topology occur before the perturbation is
switched off, and again no anomalous resistivity is excited afterwards.

Another obvious modification consists in lowering the threshold $v_{cr}$ for
the onset of anomalous resistivity.  This parameter is usually set at an
intermediate value for numerical convenience, to both enable recurrence of
anomalous resistivity and to avoid its excitation over widespread regions of
the box.  A typical value in two-dimensional simulations is $v_{cr}=3$.  From
a physical point of view, a low value of $v_{cr}$ only slightly exceeding
unity would be most reasonable for a study of spontaneous reconnection in
cosmic plasmas, since it corresponds to a configuration in which the gradient
scales in $x$ and $z$ are much larger than the current sheet width and the
onset of resistivity enhancement in one place results from small fluctuations
or a slight systematic inhomogeneity.  We have therefore integrated the system
with $l_z=l_x=l_y=0.8$ using $v_{cr}=1.1\mbox{--}1.5$ and obtained similar
dynamical behavior in this whole range, including re-excitation of anomalous
resistivity.  This counteracts the general diffusion that dominated run~1 and
enables build-up of a long-lasting spontaneous fast reconnection regime.  The
structures and dynamics in this case are only partly similar to the 2D system
and will be explored further by run~3 in subsequent sections.

Finally, we consider very small values of $\beta$ permitting higher
compressibility of the plasma.  This should render excitation of anomalous
resistivity according to Eq.~(\ref{etaandef}) easier, since a higher pile-up
of magnetic flux and current density, and stronger density reductions, are
enabled.  It turns out, however, that reducing $\beta$ below our standard
value of 0.15 leads to only a minor increase (less than one per cent) of the
maximum electron-ion drift velocity that is attained during the evolution.
Generally, the influence of the plasma beta on the evolution of the considered
systems has been found to be very weak in the range $\beta=0.0015-1.5$
(runs~6--8 in Table~\ref{Tab1}).

Summarizing this parametric search, we find that appropriate setting of the
threshold for the onset of anomalous resistivity enables the build-up of 
spontaneous reconnection in the case of an isotropic initial 
perturbation.

\subsection{Structure of the reconnection region with a central X-line}

First we discuss the case $l_z=8$ (run~2).  The three-dimensional structure of
the current sheet at times of a fully developed fast reconnection regime shows
that the dynamics is surprisingly well confined to a slab $|z|\alt2\,l_z$
around the $z=0$ plane.  This is a consequence of the inflow $u_z$, which
continues as long as the central X-line exists and inhibits spreading of the
reconnection process in $z$, of the absence of a guide field component, and of
the neglect of viscosity (which is reasonable for dilute cosmic plasmas).
Magnetic field lines outside of this slab do not show the X-type topology, as
can be seen in Fig.~\ref{B_lines} at $t=95$.  The bending of the field lines
within the current sheet ($y<1$) reflects the inflow $u_z$ along the $z$ axis
and the diversion of the outflow around the pressure maximum (``plasmoid''
structure) which is pushed outward by the main reconnection outflow $u_x$.
The outflow becomes Alfv\'enic at the rear side of the outgoing plasmoid (see
below, Fig.~\ref{prof2}).

The isosurfaces of the current density displayed in Fig.~\ref{j_surf} for
run~2 show the slab-like character of the system perhaps most clearly:  the
sheet remains practically undisturbed at $|z|\agt2\,l_z$
[Figs.~\ref{j_surf}(a) and \ref{j_surf}(b)].  Within $|z|\alt l_z/2$ and
between the pair of plasmoids ($|x|\alt15$ at $t=95$) the structure is nearly
two-dimensional.  The cut at $z=0$ reveals the current density ridges (slow
mode shocks) known from 2D reconnection.  The occurrence of enhanced
resistivity is much more localized at the origin than the regions of high
current density and does not extend along the current density ridges (see
Figs.~\ref{comp}, \ref{prof1}, and \ref{prof2}).  This is due to the
rarefaction of the plasma at the origin, which enters into the velocity-based
criterion for resistivity enhancement.  Thus, the essential features of
Petschek's reconnection model (Alfv\'enic outflows, localized resistivity
enhancement at a central X-line, and slow mode shocks) are set up for
$l_z>l_z^*$ within the nearly two-dimensional slab.

The configuration formed from an isotropic initial perturbation is more
complicated and only partly similar to the Petschek configuration.  In run~3
($l_z=0.8$, $v_{cr}=1.5$) anomalous resistivity is never excited at ${\bf
x}=0$.  Instead, oblique flows and flux pile-up by the inflow $u_z$ lead to a
symmetric double structure with maxima of $\eta$ initially lying at the $z$
axis slightly inward of the point of maximum slope of the initial resistivity
perturbation profile; the point $z_1$ of maximum $\eta$ lies in the range
$|z|\approx0.2$--0.4 for $14<t<72$.  An inflow $u_y(y>0)<0$ is built up at the
maxima of $\eta$, leading to a Petschek-like structure of reconnection in a
neighborhood of the planes $|z|=z_1$.  A magnetic X-configuration with the
corresponding $x$-$y$ flow and the classical current density ridges exists
there temporarily.  These two slabs form a sandwich-like structure about the
midplane $z=0$, in which the flow along the $y$ axis is reversed,
$u_y(y>0)>0$, and reconnection is inhibited (Fig.~\ref{sandwich}).  The
outflow $u_x$ that results from the initial perturbation and the current
density ridges extend between the slabs through the plane $z=0$.  The
available numerical resolution does not permit a clear distinction between
whether the X-line continues through $z=0$ or an O-line is formed at $|z|\ll
z_1$ already in this early phase.

The profiles along the $z$ axis of pressure, density, and velocity in
Fig.~\ref{profz} show the flow, driven by the pressure gradient
$-\partial_zp/2$, at characteristic times for runs~2 and 3.  The structures
are similar in both runs with spatial scales being of the order $l_z$ during
the Petschek-like phase.

\subsection{Secondary tearing and long-term evolution}
\label{long}

The plot of the electric field at the X-lines in Fig.~\ref{R_rate} shows the
temporal evolution of the reconnection process.  We refer to $R={\eta}j$ as
the reconnection rate in our system with $B_z(t\!=\!0)=0$, where the X-lines
are magnetic nulls.  Runs~2 and 3 and a two-dimensional comparison run are
shown.  The latter has physical and numerical parameters identical to run~2,
except for the use of a uniform grid with resolutions $\Delta x_0$ and 
$\Delta y_0$ as given in Table~\ref{Tab1} for run~2.  Again we begin with a
discussion of run~2.  The reconnection rate at the X-line formed at 
${\bf x}=0$ develops a single, well-defined peak (thick continuous lines in
Fig.~\ref{R_rate}) which is of similar magnitude and shape for $l_z=8$ and
$l_z=\infty$.  Most importantly, $R$ does not stabilize after a Petschek-like
configuration is established, but shows a monotonic decay with a decay time
comparable to the rise time.  This is in contradiction to the conclusions of
Ugai and Shimizu\cite{US96} who had found some indications in favor of
build-up of a quasi-steady reconnection regime (the rate of change of the
flows and of the peak anomalous resistivity began to decrease over a period of
$\sim10\tau_A$ at the end of their run).  Our calculation in a larger box over
a longer period shows that, in fact, a quasi-steady state is not reached.
Ugai and Shimizu have argued that a balance between convection of field lines
by the inflow $u_y$ and magnetic diffusion is reached at the central X-line
because $\eta=\eta(j/\rho)$ is permitted to rise freely, which enables a
(quasi-) steady state of reconnection in 3D as well as in 2D.  Their argument
implicitly relies on the reconnection geometry in the $x$-$y$-plane being
stationary.  However, the structure of the current sheet in the neighborhood
of the X-line evolves during the build-up of the reconnection regime.  It is
the evolution of the volume where anomalous resistivity is excited which leads
to a decrease of the reconnection rate at ${\bf x}=0$ in the long-term
evolution.

Figure~\ref{prof1} shows profiles along the $x$ axis of $j$, $\rho$, and the
resulting anomalous resistivity $\eta_{an}$, for run~2.  The regions of
enhanced $j(x)$ and $\eta_{an}(x)$ at the origin broaden with time
considerably.  At the same time, the width of the corresponding profiles along
the $y$ axis stays roughly constant, due to the inflow $u_y$.  With anomalous
resistivity extending to $|y|\approx0.2$, the aspect ratio of the $\eta_{an}$
area has reached a value $\delta x_{\eta}/\delta y_{\eta}=33$ by $t=127$.
Essentially, the inner section of the current sheet has undergone a transition
from a Petschek configuration to a Sweet-Parker configuration.  Consequently,
the reconnection of field lines and the level of anomalous resistivity drop,
primarily at the origin (where the pull by the outflow $u_x$ becomes weaker
than at the ends of the extended $\eta_{an}$ area).  Two symmetric maxima of
the $\eta_{an}$ profile remain, which immediately lead to the formation of a
pair of new magnetic X-lines at $t\approx150$.  An inflow along the $z$
direction is created also at the new X-lines, but it has only a weak effect
because the reconnection inflow $u_y$ is already present at a larger scale.
The outflow from the new X-lines drives a transition of topology at the
origin, where an O-line is created.  This change of topology is enabled by the
still existing (although rapidly decaying) anomalous resistivity at ${\bf
x}=0$.  The reconnection rate in run~2 drops from $R=1.6\times10^{-2}$ at
$t=95$ to $R=8.4\times10^{-3}$ at $t=152$.  The Sweet-Parker reconnection
rate, based on $\delta x_{\eta}$ and the average value
$\langle\eta_{an}\rangle_{\delta x_{\eta}}$, is found to drop similarly,
$R_{SP}(t\!=\!95)=2.9\times10^{-2}$ and $R_{SP}(t\!=\!152)=1.6\times10^{-2}$,
thus supporting our interpretation.  In comparison, the Petschek reconnection
rate, based on the $x$ extent of the Petschek-like configuration (i.e., the
$x$ extent of the current density ridges behind the outgoing plasmoid) and
$\langle\eta_{an}\rangle_{\delta x_{\eta}}$, stays approximately constant,
$R_{P}(t\!=\!95)=4.4\times10^{-2}$ and $R_{P}(t\!=\!152)=4.1\times10^{-2}$.

The creation of the pair of new X-lines has been found in 2D simulations of
spontaneous reconnection and termed secondary tearing.\cite{SR87} The
evolution of the 3D system investigated here is similar to the 2D results.  We
remark that the inflow $u_z$, which peaks at the $z$ axis, supports the
topology change at the origin by its tendency to revert the flow $u_y$.  This
is particularly true for small values of $l_z$ (see below).  It has been
argued\cite{U92} that secondary tearing only occurs if the model for
$\eta_{an}$ is based on the current density and does not occur if the model is
based on the electron-ion drift velocity.  In the latter case, the density
minimum at the origin was supposed to lead to a permanent localization of the
anomalous resistivity.  However, with the increasing extent of the anomalous
resistivity area, the density minimum becomes progressively flatter, and the
evolution is similar, only somewhat delayed, for the drift velocity criterion.
Secondary tearing occurred in all runs which exhibited spontaneous
re-excitation of anomalous resistivity (for various parameter settings in
addition to Table~\ref{Tab1}) at late stages ($t\agt150$).

The plot of the field lines and of the profiles $u_x(x)$ and $p(x)$ for run~2
in Fig.~\ref{prof2} demonstrates that the secondary tearing is not caused by
the open boundary in the outflow direction ($|x|=L_x$):  it occurs at a time
when the disturbance created by the initial resistivity perturbation has not
yet reached that boundary.  It is also not caused by the nonuniformity of the
grid, since the new X-line is formed at $x\approx5$ while the grid is uniform
for $|x|<10$ (Table~\ref{Tab1}).

The topology change connected with the secondary tearing does not inhibit fast
magnetic reconnection, instead it leads to re-amplification of the process.
The reconnection rate remains variable because the plasmoid in the center of
the system, which is permanently driven by the outflow from the new X-lines,
executes oscillatory motions (it may even tear and merge again, as indicated
in Fig.~\ref{prof2} at $t=190$ and by the small peak in the second panel of
Fig.~\ref{R_rate} at the same time) and because the $\eta_{an}$ area continues
to show a slight tendency to elongate outwards in a time-variable manner.
Both effects lead to variations of $\eta_{an}$ at the new X-lines.  Repeated
secondary tearing of this elongated current sheet section, typical of the 2D
system (see the third panel of Fig.~\ref{R_rate} and Ref.~\onlinecite{SR87}),
occurs only once ($t=500\mbox{--}540$) and is of minor dynamical importance.
Obviously this effect requires even larger $l_z$ because the tearing mode is
most easily excited as a 2D instability\cite{SS97} (although its nonlinear
development may be three-dimensional\cite{DAZ92}).  Increasing numerical
diffusion in the nonuniform part of the grid ($|x|>10$) may also have an
influence.

A plasmoid with continuously rising pressure is formed in the central O-line
region by the outflow from the adjacent X-lines.  This plasmoid relaxes
primarily by setting up a pair of outflow jets along the $z$ axis.  These
outflow jets, which are strongly confined (roughly to the cross section of the
plasmoid), are shown for run~2 in Fig.~\ref{jets}(a).  Their velocity reaches
$0.7\,V_A$ during $t=200\mbox{--}380$ (Fig.~\ref{profz}).  In 2D the
relaxation is only possible by slowly shifting the new X-lines outward along
the $x$ axis.\cite{SR87} Also the area of anomalous resistivity starts to grow
in the $z$ direction, forming a pair of ropes at the surface of the plasmoid
(Fig.~\ref{etaropes}).  At the same time, the reconnection rate at the new
X-lines decreases at the $x$ axis, since an increasing amount of field line
bending is required to convect flux toward the new X-lines, which are staying
close to the swelling plasmoid, and since the increasing pressure in the
plasmoid resists the reconnection outflow from the new X-lines.  This results
in a splitting and displacement of the maxima of $\eta_{an}$ and $R$ off the
$x$ axis along these ropes toward large values of $|z|$ after $t=450$.  The
second panel of Fig.~\ref{R_rate} displays the decrease of the reconnection
rate at $z=0$ as a continuous line and the global maximum of the reconnection
rate as a dashed line.  The latter is observed to stay almost constant in the
short interval during which the maximum is displaced at high apparent velocity
to the boundary $L_z$.  The secondary tearing thus enables spreading of the
reconnection process in the $z$ direction.  However, the current limitation of
grid sizes prevents a systematic investigation of this phase.

The time evolution of the reconnected flux in the first quadrant of the
current sheet plane, 
$\Phi=\int B_y(x,0,z,t){\rm d}x{\rm d}z/\delta z_\eta(t)$, 
which is plotted in the bottom panel of Fig.~\ref{R_rate}, also indicates
enhanced reconnection after secondary tearing and the spreading of the
reconnection process in the $z$ direction.  The ``outer'' reconnected flux
($\Phi_o<0$) between the dominant X-line and the boundary $x=L_x$ and the
``inner'' reconnected flux ($\Phi_i>0$), which accumulates between the
dominant X-line and the $z$ axis after secondary tearing, are plotted
separately.  One has to keep in mind, however, that these quantities still
carry signatures of several processes, which are difficult to disentangle and
quantify.  Both parts of the flux are influenced by convection across the
boundaries $x=L_x$ or $z=L_z$.  The area of integration, and consequently the
proportion between $\Phi_o$ and $\Phi_i$, depend on the location of the
dominant X-line, which is strongly shifted in the 2D run.  Coalescence of
magnetic islands after secondary tearing annihilates previously reconnected
flux.  Finally, there is some arbitrariness in the normalization of $\Phi$ for
the 3D runs, required for the comparison with the 2D run:  it can be done
using $l_z$, $L_z$, or the instantaneous extent of the anomalous resistivity
region $\delta z_\eta(t)$.  Nevertheless, one can see that (1) maxima of 
${\rm d}|\Phi|/{\rm d}t$ correspond to enhanced reconnection rates in the
upper panels of the figure, (2) run~2 and the 2D run lead to comparable
amounts of reconnected flux, (3) the flux $-\Phi_o$ peaks roughly at the time
at which the rear side of the plasmoid that is created by the initial
perturbation has reached the boundary $L_x$.

Turning now to the case of isotropic initial perturbation (run~3 with
$l_z=0.8$), we find that the evolution is largely, but not fully, analogous to
the case of anisotropic initial perturbation (run~2).  Again, the
Petschek-like configuration initially formed in the two symmetric slabs around
$|z|=z_1$ turns out to be not a stable configuration.  Figure~\ref{eta_evo}
shows the evolution of the anomalous resistivity in the current sheet plane
$y=0$, where the maximum of $\eta$ remains during the whole run.  The initial
elongation of the $\eta_{an}$ area in the direction of the outflow and the
subsequent decrease of $\eta_{an}(x=0)$ are similar to run~2.  The combined
effect of the coherent outflow $u_x$, which extends over a larger $z$ interval
than the anomalous resistivity and includes the plane $z=0$ (cf.\ 
Fig.~\ref{sandwich}), and the continuing inflow $u_z$ let the $\eta_{an}$ area
then approach the $x$ axis.  Secondary tearing in the plane $z=0$ occurs
immediately ($t=72$), also being supported by the still reversed flow
$u_y(0,y>0,0)>0$.  A magnetic configuration identical to the one shown in the
bottom panels of Fig.~\ref{prof2} is formed.  The subsequent evolution is
qualitatively similar to the case $l_z=8$, although the reconnection rate
remains much smaller and is far less variable.  This reflects the fact that
here the driving reconnection at the new X-line is far less coherent in $z$
and the flows $u_{x,y,z}$ have to be partly reorganized in space.  The
reconnection rate was found to depend only weakly on the parameter $v_{cr}$ in
the range $1.1\le v_{cr}\le1.5$.  The outflow jets along the $z$ axis are also
formed [Figs.~\ref{profz} and \ref{jets}(b)].  The maximum outflow velocities
are $u_x=0.8$ at $t\sim400$ and $u_z=0.3$ at $t=300\mbox{--}450$.  The
splitting of the maximum resistivity in the $z$ direction and the subsequent
rapid displacement of the maximum along the new X-lines to the boundary
$|z|=L_z$ occurs at $t>310$.  As in the case $l_z=8$, there is a strong
tendency toward a quasi-2D configuration in the numerical box, based on the
formation of $\eta_{an}$ ropes along the new X-lines (Fig.~\ref{eta_evo}).  In
a larger system, it can be expected that the reconnection rate and the outflow
velocities rise further as soon as the quasi-2D configuration extends over a
sufficiently large interval $\Delta z\gg1$ and the overall configuration
becomes similar to run~2.

The secondary tearing is essential to reach a significant reconnection rate in
the case of an isotropic initial perturbation, because it leads to growth of
the area of anomalous resistivity in the $z$ direction.

\subsection{Parallel current density and electric field}

Field lines of ${\bf{j}}({\bf{x}})$ are displayed for run~2 in
Fig.~\ref{j_lines}.  These visualize the current diversion which is caused by
the finite $z$-extent of both the resistivity region and the reconnection
flows, $u_x$ and $u_y$.  Lines of current flow starting close to the current
sheet plane $y=0$ are diverted mainly in the $x$ direction, while lines
starting at a certain distance from $y=0$ are diverted mainly in the $y$
direction, to form the current density ridges shown in Fig.~\ref{j_surf}.  The
current diversion is related to the generation of current components parallel
to the magnetic field, which are of particular significance because the
resulting parallel electric field, $E_{||}=\eta_{an}j_{||}$, may dominate the
acceleration of particles and because the currents may be directly observable
as the ``substorm current wedge'' in events of magnetospheric activity.  Note
that $E_{||}=j_{||}=0$ in the 2D evolution of neutral current sheets.  The
generation of parallel current components is caused by the plasma flow as well
as by the existence of nonuniform (anomalous) resistivity. This can be seen by 
considering the current helicity density \cite{BER84,SEE96}
\begin{eqnarray*} 
h_c={\bf j}\cdot{\bf B} \,.
\end{eqnarray*}
Its time derivative is related to a temporal change of the parallel current
density. In dimensionless units: 
\begin{eqnarray} 
\frac{\partial h_c}{\partial t}
                   &=&\partial_t{\bf j}\cdot{\bf B}+
                      {\bf j}\cdot\partial_t{\bf B}\nonumber\\
                   &=&[{\boldmath\nabla\times}
                       {\boldmath\nabla\times}
                      ({\bf u}{\boldmath\times}{\bf B})]\cdot{\bf B}
                                                                \nonumber\\
                   & &-\,[{\boldmath\nabla\times}
                       {\boldmath\nabla\times}
                      (\eta{\bf j})]                    \cdot{\bf B}
                                                                 \nonumber\\
                   & &+\,[{\boldmath\nabla\times}
                      ({\bf u}{\boldmath\times}{\bf B})+
                       \eta\,\Delta{\bf B}]\cdot{\bf j}\,.
\label{gradeta}
\end{eqnarray}
The second term shows that contributions to the growth of $j_{||}$ result not
only from the presence of resistivity but also from nonvanishing first and
second spatial derivatives.  The derivatives of $\eta$ are particularly
important at large Lundquist numbers, where the terms proportional to $\eta$
become small.  That these contributions are indeed essential can be seen in
Fig.~\ref{jpar&eta}, where contours of $j_{||}$ and $\eta$ are plotted for
run~2 ($l_z=8$) during the Petschek-like reconnection phase ($t=95$) and after
secondary tearing has occurred ($t=190$).  At both times the highest parallel
current density components are created close to the surface of the region of
nonvanishing resistivity, where the first and second derivatives of $\eta$ are
both large.  Figure~\ref{j_||}(a) displays isosurfaces of $|j_{||}|$ for the
same dataset ($t=190$) at two levels.  The concentration of the highest
parallel current densities close to the surface of the resistivity region is
again apparent.  The lower-level isosurface shows that the overall structure
of the $j_{||}$ region follows the current density ridges seen in
Fig.~\ref{j_surf} and that there is a second region of enhanced parallel
current components at the rear side of the outgoing plasmoid.  Significant
$j_{||}$ components are thus also created at lines of current flow far away
from the resistivity region (cf.  Fig.~\ref{j_lines}).  The latter effect is
due to the shear of the outflow, which is diverted around the outgoing
plasmoid, and has been discussed in detail
previously.\cite{SCH91A,SCH91B,BH96} The highest parallel current densities
occur as a result of secondary tearing ($t=180$--420 in run~2 and $t=350$--450
in run~3) at the surface of the central plasmoid close to the area of
anomalous resistivity [Fig.~\ref{jpar&eta} and Fig.~\ref{j_||}(b)].  The
maximum parallel current densities reached in run~2 [$\max(j_{||})=2.4\,j(0)$]
and run~3 [$\max(j_{||})=2.1\,j(0)$] are comparable.

The parallel electric field peaks where the volumes of anomalous resistivity
and high parallel current density intersect, i.e., at the surface of the
central plasmoid close to the new X-lines after secondary tearing
(Fig.~\ref{jpar&eta}).  For both runs~2 and 3, the temporal maximum of
$E_{||}$ is related to a strong oscillation of the central plasmoid, which
tends to tear in the $x$ direction but merges again,
$\max[E_{||}(x=4.0,y=0.09,z=3.5,t=197)]=1.3\times10^{-3}$ in run~2 and
$\max[E_{||}(2.7,0.7,0.6,352)]=1.7\times10^{-3}$ in run~3.

\section{CONCLUSIONS AND DISCUSSION}
\label{conclusions}

In this paper we have investigated, within the framework of one-fluid MHD, the
three-dimensional dynamics of magnetic reconnection in a compressible Harris
current sheet with initially antiparallel magnetic field and dynamically
coupled resistivity.  The three-dimensional development was initiated by
applying a resistivity perturbation bounded in $x$, $y$, and $z$.  Particular
attention was paid to the question of the development of the so-called
spontaneous fast reconnection regime.  The results can be summarized as
follows.

(1) The dynamical development depends strongly on the elongation $l_z$ of the
initial perturbation in the $z$ direction relative to its dimensions $l_x$ and
$l_y$.  Only for sufficiently anisotropic resistivity profiles with
$l_z>l_z^*>l_{x,y}$ will the inflow along the $z$ axis be sufficiently weak
so that the reconnection inflow in the $y$ direction and a fast reconnection
regime can develop in a manner similar to the two-dimensional case.  For the
Harris sheet with $l_x\!\sim\!l_y\!\sim\!1$ and $v_{cr}=3$, the critical
elongation $l_z^*$ lies in the range 4--8.  Over a period of
$\sim10^2\,\tau_A$, a Petschek-like reconnection regime is built up but
remains bounded to a slab around the plane $z=0$, the width of which is
approximately equal to twice the $z$ extent of the initial perturbation.

(2) The system is not able to sustain the Petschek reconnection regime in a
(quasi-) steady manner, instead the region of anomalous resistivity at the
X-line becomes increasingly elongated with time in the $x$ direction, thus
temporarily resembling a Sweet-Parker configuration.  This leads to a decrease
of the reconnection rate in the region of the initial perturbation.  As a
result, the system shows a transition of topology known as secondary tearing,
i.e., a symmetrical pair of X-lines is formed and the original X-line is
replaced by an O-line.  This process is supported in the three-dimensional
case by the inflow along the $z$ axis and occurs even with the drift
velocity-based criterion for resistivity enhancement.  The secondary tearing
leads to enhanced and time-variable reconnection, to a second pair of outflow
jets directed along the O-line, and to expansion of the reconnection 
process along the newly formed X-lines in the $z$ direction. 

(3) The requirement on the anisotropy of the initial perturbation can be
relaxed by lowering the threshold for the excitation of anomalous resistivity
substantially below the value commonly used in two-dimensional simulations.
The system with an isotropic initial perturbation then evolves through a
partly Petschek-like reconnection regime but experiences secondary tearing
relatively early.  Subsequently, the evolution proceeds in a qualitatively
similar manner as in the case of a strongly anisotropic initial perturbation.
With the expansion of the reconnection process along the new X-lines,
significant reconnection rates are achieved after several $10^2$ Alfv\'en
times.

(4) The three-dimensional reconnection process forms substantial current
density components along the magnetic field, 
$\max[j_{||}({\bf x},t)]\approx2.4\,j(0)$ for $l_z=8$ and $\approx2.1\,j(0)$
for $l_z=0.8$.  Their maximum lies at the surface of the plasmoid formed by
secondary tearing near the boundary of the anomalous resistivity region (i.e.,
near the new X-lines).  Also the maximum parallel electric field occurs near
the new X-lines, within the region of anomalous resistivity.  It is of order
$E_{||}\sim(1\mbox{--}2)\times10^{-3}$ and is rather insensitive to the
elongation $l_z$ of the initial perturbation.

The choice of a strongly anisotropic perturbation ($l_z\gg l_x$ in run~2) has
permitted a detailed comparison of the different phases of three-dimensional
spontaneous resistive reconnection in initially antiparallel magnetic fields
(neutral current sheets) with the two-dimensional case.  However, as discussed
already in the introduction, this configuration with the anisotropy oriented
across the magnetic field is not expected to occur spontaneously (caused,
e.g., by fluctuations in a marginally stable equilibrium) in cosmic plasmas.
Possibly an ideal MHD instability can be found which enforces such a
configuration.  Simulations of the reconfiguration of an arcade of magnetic
loops in the solar atmosphere\cite{AM96} are suggestive in this direction and
have been taken in the literature to justify two-dimensional models of the
subsequent reconnection, but this particular instability requires a large
guide field component.

Our study of spontaneous reconnection initiated by an isotropic perturbation
($l_x=l_y=l_z$ in run~3) required that a lower threshold for the onset of
anomalous resistivity be chosen than commonly used in 2D simulations, which,
however, appears physically reasonable.  Although this system did not show all
the attributes of ``fast Petschek-like reconnection,'' substantial
reconnection rates were achieved at least over several $10^2\,\tau_A$ and
clear indications have been obtained that the reconnection process also
spreads in the $z$ direction to become a large-scale disturbance.  This
suggests that three-dimensional fast reconnection may occur spontaneously and
lead to macroscopic energy release events in neutral current sheets.

As a further possibility to trigger long-lasting, large-scale, and possibly
fast magnetic reconnection by a three-dimensionally localized initial
perturbation, one has to consider a sheared current sheet with a nonvanishing
guide field ($B_z(t\!=\!0)\ne0$).  In that case an anisotropy $l_z\gg l_x$ of
the resistivity perturbation would be oriented {\em along} the magnetic field
in the center of the current sheet ($y=0$) and can therefore be expected to
occur spontaneously.  Three-dimensional magnetic reconnection in such a
sheared field configuration, where the midplane reflection symmetry is broken
and the dynamics is strongly modified in comparison to the system considered
here, will be investigated in a future paper.

Finally, a long-lasting and possibly steady regime of three-dimensional fast
reconnection may be obtained by prescribing an inflow toward the
sheet,\cite{SW84} similar to two-dimensional simulations.\cite{AJ96} The
inflow $u_y(|y|\to\infty)$ may be relatively weak, since rather low values of
$u_y$ [$\sim(1\mbox{--}3)\times10^{-2}$] occurred in our simulations even at
times of peak reconnection rates.  Such driven reconnection is expected to
occur at the dayside of the magnetosphere and possibly at the boundary of
newly emerging magnetic flux in coronae.

The long-term evolution of three-dimensional spontaneous reconnection has been
investigated here under the constraint of symmetry with respect to the planes
$x=0$, $y=0$, and $z=0$, dictated by practical limitations of
three-dimensional grid sizes.  Strong symmetric outflows $u_x$ and $u_z$ have
been found.  It is conceivable that asymmetries with respect to $x=0$ or $z=0$
would lead to predominantly one-sided outflows $u_x$ or $u_z$, respectively,
and to a different long-term evolution.  For weak asymmetries the ejection of
the plasmoid formed through secondary tearing may occur.  This effect has been
seen in two-dimensional simulations\cite{U85,SK96} and will be studied in 3D
as well.

The growth of the region of anomalous resistivity into the direction of the
reconnection outflow $u_x$ turned out to be an essential element of the
dynamics that leads to secondary tearing and subsequent growth of the
perturbation into the $z$ direction.  This effect is related to the inability
of the fluid (at least in standard MHD) to carry a sufficient amount of
magnetic flux into a strongly localized diffusion region to support the
Alfv\'enic outflow of reconnected flux in steady state.\cite{KU98} Higher
inflow velocities into the diffusion region occur in more general fluid or
kinetic treatments of collisionless and resistive
reconnection,\cite{BSD97,HE99} where separate electron and ion scales form.
How these faster flows match to the outer regions, where the electrons and
ions are coupled, and thus the question of whether the growth of the diffusion
region is enforced by the large-scale flow pattern irrespective of its
microscopic physics, requires further investigation.

\acknowledgments
We thank G. T. Birk, J. Dreher, K. Schindler, and M. Scholer for helpful
discussions. The paper benefitted also from critical comments by the referee. 
This work was supported by Grants No. 50QL9208, No. 50QL9301, and 
No. 50OC9706 of the Deutsche Agentur f\"ur Raumfahrtangelegenheiten and Grant 
No. 28-3381-1/57/97 of the Ministerium f\"ur Wissenschaft und Forschung 
Brandenburg. The John von Neumann-Institut f\"ur Computing, J\"ulich 
granted Cray T90 computer time. 

\clearpage


\clearpage

\begin{table}
\caption{Parameters of the three-dimensional simulation runs. The resolutions 
$\Delta x_0$, $\Delta y_0$, $\Delta z_0$ refer to the inner uniform part of 
the nonequidistant grid, whose size is given by $L_x'$, $L_y'$, and $L_z'$.} 
\begin{tabular}{ccccccccccccccc}
Run&$l_x$&$l_z$&$v_{cr}$&$\beta$&$L_x$&$L_y$&$L_z$&$\Delta x_0$
                                                         &$\Delta y_0$
                                                           &$\Delta z_0$
                          &$L_x'$&$L_y'$&$L_z'$&$\eta_{an}(t\!>\!t_0)$    \\
\hline
1  & 0.8 &  1  &   3    & 0.15  &  20 &  4  &  25 & 0.10 & 0.045 & 0.075 
                          &   3 &  1  &   2 &  no                    \\ 
2  & 0.8 &  8  &   3    & 0.15  &  90 &  4  &  40 & 0.10 & 0.045 & 0.25  
                          &  10 &  1  &   3 &  yes                   \\ 
3  & 0.8 & 0.8 &  1.5   & 0.15  &  90 &  4  &  10 & 0.10 & 0.045 & 0.025 
                          &  10 &  1  &  1.3&  yes                   \\ 
4  & 4.0 &  1  &   3    & 0.15  &  40 &  4  &  20 & 0.10 & 0.045 & 0.1 
                          &   3 &  2  &   3 &  no                    \\ 
5  & 16.0&  1  &   3    & 0.15  &  40 &  4  &  20 & 0.10 & 0.045 & 0.1 
                          &   3 &  2  &   3 &  no                    \\ 
6  & 0.8 &  1  &   3    & 0.015 &  20 &  4  &  25 & 0.10 & 0.045 & 0.075 
                          &   3 &  1  &   2 &  no                    \\ 
7  & 0.8 &  1  &   3    & 0.0015&  20 &  4  &  25 & 0.10 & 0.045 & 0.075 
                          &   3 &  1  &   2 &  no                    \\ 
8  & 0.8 &  1  &   3    & 1.5   &  20 &  4  &  25 & 0.10 & 0.045 & 0.075 
                          &   3 &  1  &   2 &  no                       
\end{tabular}
\label{Tab1}
\end{table}

\begin{figure}
\begin{center}
\epsfig{file=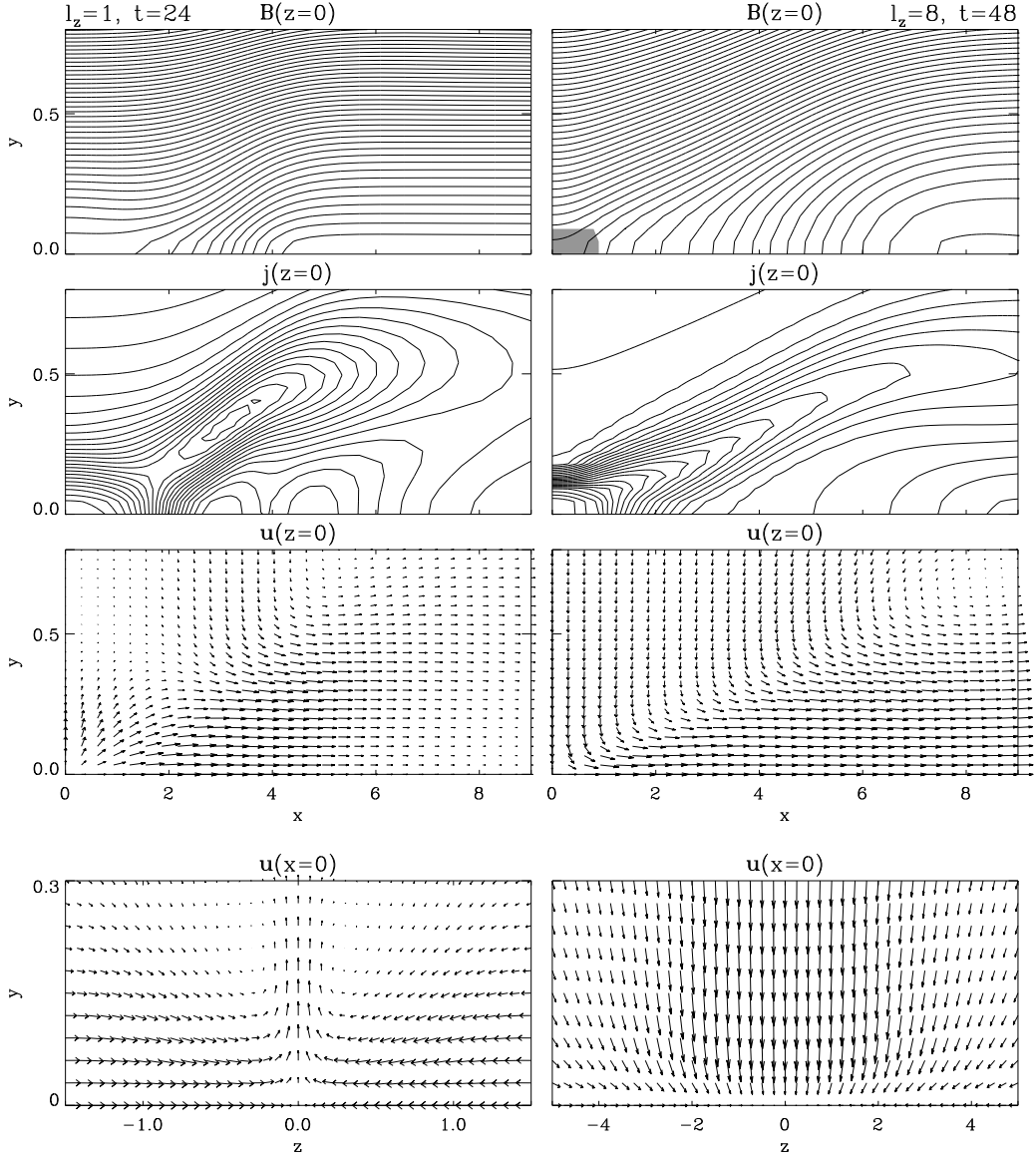}
\end{center}
\caption{Magnetic field, current density, and velocity in the plane $z=0$, 
         and velocity in the plane $x=0$ for run~1 ($l_z=1$) at $t=24$ (left
         panels) and for run~2 ($l_z=8$) at $t=48$ (right panels). The 
         maximum outflow velocities are $\max(u_x)=0.23$ for both runs. The
         gray shaded area is the region where anomalous resistivity is 
         excited. The peak values of the velocity components in the bottom
         panels are $|u_y|\approx0.02$ and $u_z\approx0.1$ for both runs. 
         (Peak velocities for run~2 at $t=24$ are about half the values 
         given here.) Peak current densities are $\max(j)=1.7$ (run~1)
         and $\max(j)=3.0$ (run~2).}
\label{comp}
\end{figure}
\begin{figure}
\begin{center}
\epsfig{file=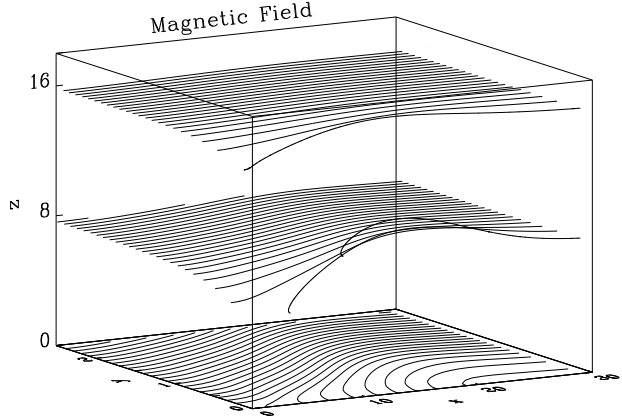}
\end{center}
\caption{Magnetic field lines passing through $x=30$ and $z=0,\,l_z,\,2l_z$,
         respectively, for run~2 ($l_z=8$) at $t=95$.}
\label{B_lines}
\end{figure}
\begin{figure}
\caption{Surfaces of constant current density $j=|{\bf{j}}|$ for run~2
         ($l_z=8$). 
         (a) The 20 per cent level of $\max(j)=4.0$ (located at the origin) 
             at $t\!=\!95$.
         (b) The 10 per cent level of $\max(j)=5.8$ (located at the origin) 
             at $t\!=\!190$ (before plasmoid ejection). 
         (c) The 10 per cent level of $\max(j)=4.9$ located at the new
             X-lines at $t\!=\!354$ (after plasmoid ejection).}
\label{j_surf}
\end{figure}
\begin{figure}
\begin{center}
\epsfig{file=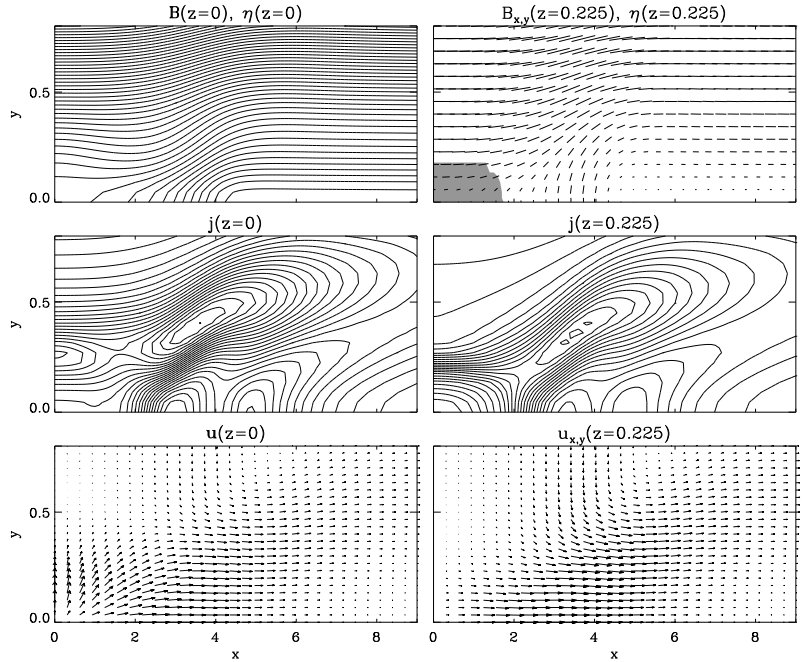,height=9cm}
\end{center}
\caption{Magnetic field, current density, and velocity for run~3 ($l_z=0.8$) 
         at $t=24$ in the plane $z=0$ (left panels) and in the plane 
         $z=z_1=0.225$, which contains the maximum of $\eta_{an}$ at this
         time (right panels). Anomalous resistivity (gray shaded) is excited
         only for $z\ne0$ at this time. Peak velocities are
         $u_x(z\!=\!0)=0.25$, $u_y(z\!=\!0)=0.034$, $u_x(z\!=\!z_1)=0.22$, 
         $u_y(z\!=\!z_1)=-0.01$. Peak current densities are
         $\max[j(z\!=\!0)]=1.3$ and $\max[j(z\!=\!z_1)]=1.8$.}
\label{sandwich}
\end{figure}
\begin{figure}
\begin{center}
\epsfig{file=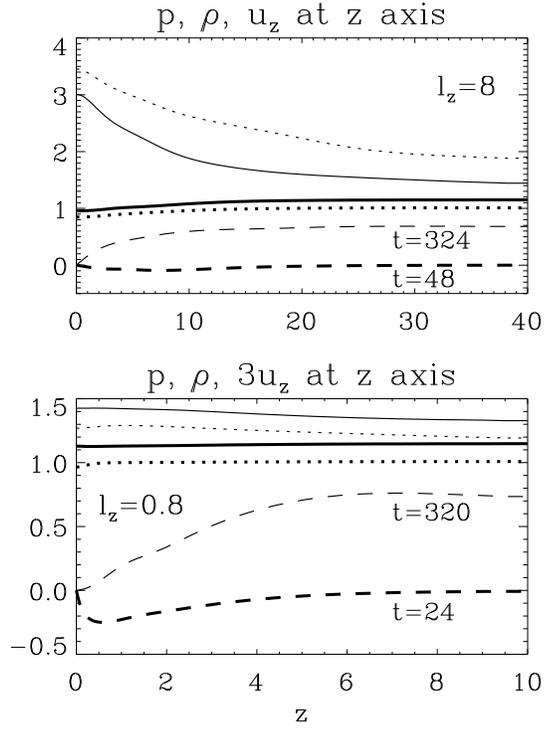}
\end{center}
\caption{Profiles of the pressure (solid lines), density (dotted), and 
         velocity (dashed) for runs~2 (top panel) and 3 (bottom panel) at a
         time of Petschek-like reconnection (thick lines) and after 
         secondary tearing (thin lines).}
\label{profz}
\end{figure}
\begin{figure}  
\begin{center}
\epsfig{file=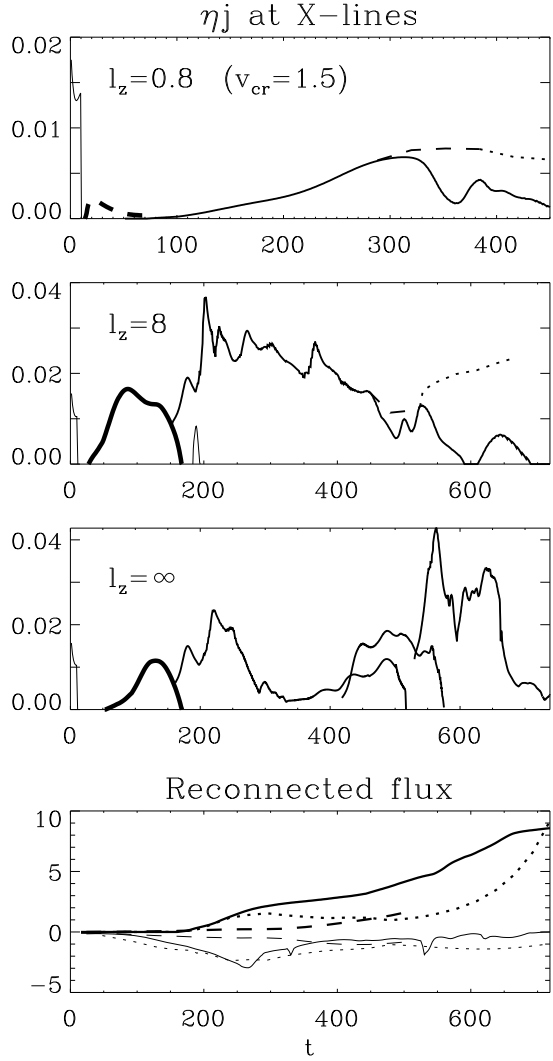}
\end{center}
\caption{Reconnection rates versus time for run~3 (top), run~2 (second panel), 
         and the 2D comparison run (third panel). Rates at different X-lines
         are shown separately. Solid lines show reconnection rates at the $x$ 
         axis, dashed lines show maximum reconnection rates occurring off 
         the $x$ axis but within the box, dotted lines refer to maximum
         reconnection rates occurring at the boundary $|z|=L_z$. Thick lines
         refer to reconnection with a central X-line (Petschek-like), 
         an intermediate line width is used for reconnection rates after 
         secondary tearing. The initial perturbation is included as a thin
         line. The narrow and weak secondary peak in run~2 at $t\approx190$ 
         results from magnetic island coalescence at ${\bf x}=0$. 
         Bottom panel: reconnected flux $\Phi$ for $l_z=\infty$ (solid 
         lines), $l_z=8$ (dotted), and $l_z=0.8$ (dashed). Thin lines
         ($\Phi<0$) refer to the reconnected flux outward of the dominant
         X-line, thick lines ($\Phi>0$) refer to the reconnected flux between
         the dominant X-line and the $z$ axis after secondary tearing.}
\label{R_rate}
\end{figure}
\begin{figure}
\begin{center}
\epsfig{file=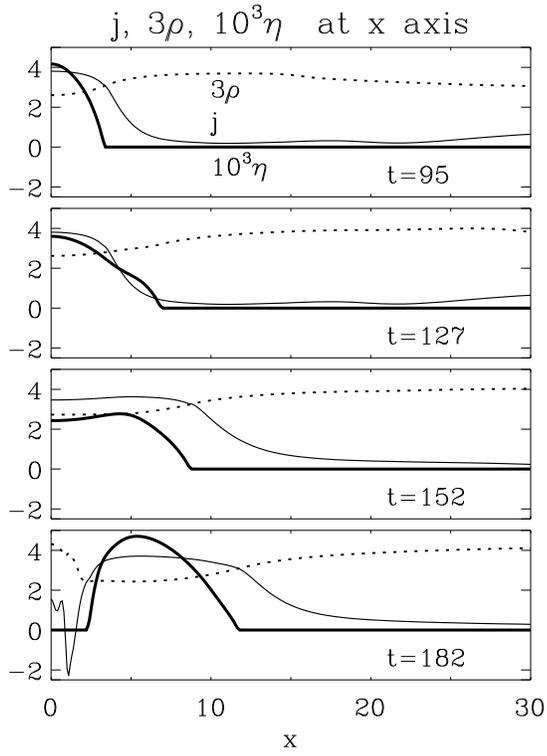}
\end{center}
\caption{Profiles of resistivity $\eta$ (thick line), current density $j$ 
         (thin line), and mass density $\rho$ (dotted) along the $x$ axis 
         for run~2 at different times.}
\label{prof1}
\end{figure}
\begin{figure}  
\begin{center}
\epsfig{file=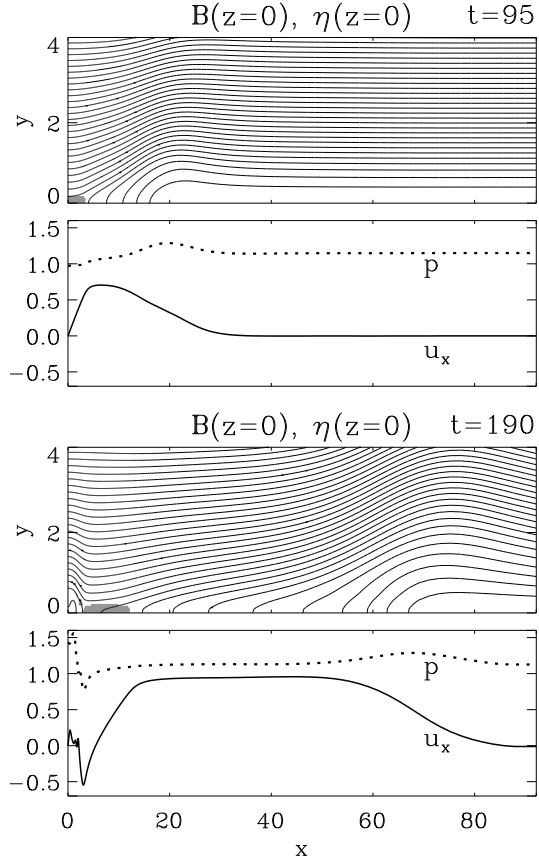}
\end{center}
\caption{Magnetic field lines in the plane $z\!=\!0$ and the profiles
         $p(x,0,0)$ and $u_x(x,0,0)$ for run~2 at $t\!=\!95$ (top panels)
         and at $t\!=\!190$ (bottom panels). The region of anomalous 
         resistivity is shown shaded.}
\label{prof2}
\end{figure}
\begin{figure}  
\caption{(a) Isosurface $|{\bf u}|\!=\!0.5$ in run~2 at $t=240$ showing 
             pairs of outflow jets along the $x$ and $z$ axes. 
         (b) Isosurface $|{\bf u}|\!=\!0.3$ for run~3 at $t=320$ showing
             similar jets.}
\label{jets}
\end{figure}
\begin{figure}  
\caption{Volume of anomalous resistivity delineating the magnetic X-lines in
         run~2 at $t=480$, where maximum reconnection occurs at $|z|=2.75$.}
\label{etaropes}
\end{figure}
\begin{figure}  
\begin{center}
\epsfig{file=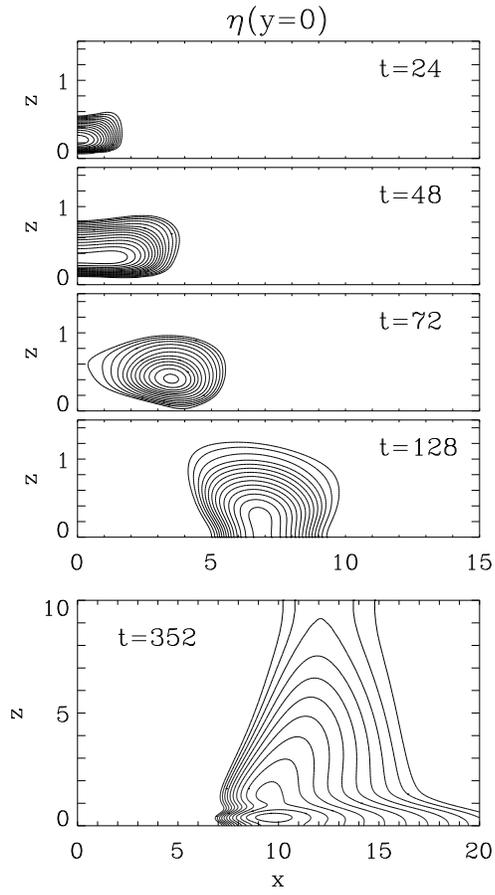,height=12cm}
\end{center}
\caption{Evolution of anomalous resistivity in the current sheet plane for 
         run~3 ($l_z=0.8$).}
\label{eta_evo}
\end{figure}
\begin{figure}  
\begin{center}
\epsfig{file=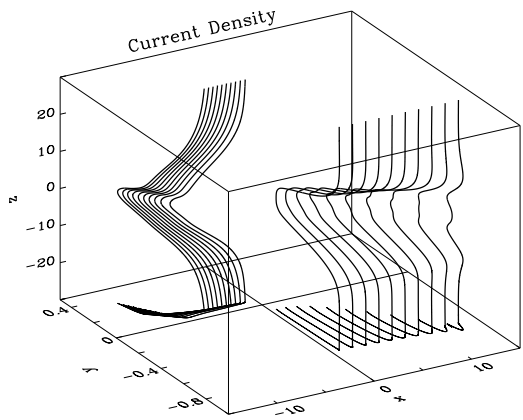,height=6cm}
\end{center}
\caption{Lines of current flow (thick lines) and their projections onto the
         $x$-$y$-plane (thin lines) for run~2 ($l_z=8$) at $t=95$. The lines
         start either at $y=0.07$ or at $y=-0.6$ at the bottom of the box.}
\label{j_lines}
\end{figure}
\begin{figure}  
\begin{center}
\epsfig{file=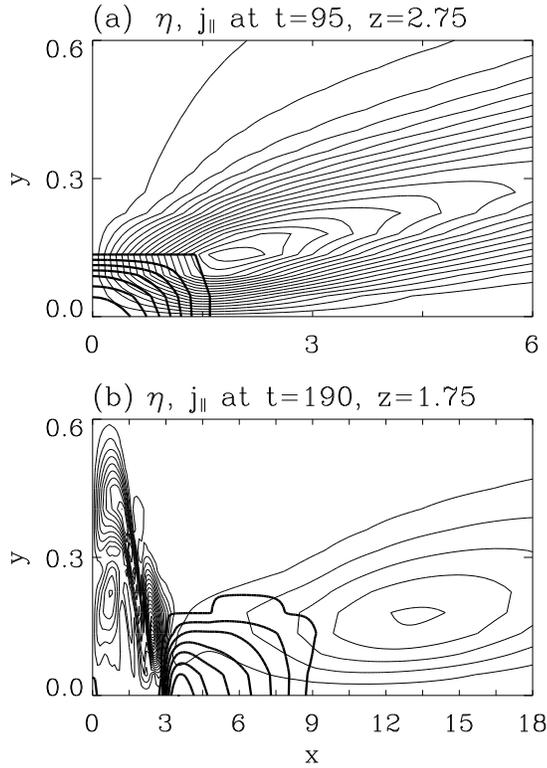}
\end{center}
\caption{Contours of $|j_{\parallel}|$ (thin lines) and $\eta_{an}$ (thick
         lines) for run~2 in the planes containing the maximum
         $|j_{\parallel}|$ during the Petschek-like phase ($t=95$) and after
         secondary tearing ($t=190$). Run~3 yields a similar picture (not
         shown).}
\label{jpar&eta}
\end{figure}
\begin{figure}
\caption{(a) Isosurface of $|j_{\parallel}|$ for run~2 at $t=190$. The 14 
             per cent level is displayed in the left half of the box ($x<0$),
             and the 7 per cent level is displayed in the right half ($x>0$).
             The maximum parallel current density is
             $j_{\parallel}(x=2.4,y=0.13,z=1.75,t=190)=1.4\,j(0)$. 
         (b) The same for run~3 at $t=384$ [4 per cent of
             $\max(j_{\parallel})=j_{\parallel}(1.7,0.8,0.2,384)=2.1\,j(0)$].}
\label{j_||}
\end{figure}

 \end{document}